\documentclass[aps,amsmath,amssymb,prl,nofootinbib,twocolumn]{revtex4}
\usepackage{amssymb,graphics,graphicx,epstopdf,color,slashed}

\begin{document}

\title{\Large Comment on ``Could the Excess Seen at $124-126$ GeV Be due to the Randall-Sundrum Radion?''}

\author{ Yong Tang
}
\affiliation{Physics Division, National Center for Theoretical Sciences, Hsinchu, Taiwan}


\maketitle

The excess seen at $125$ GeV at both ATLAS \cite{ATLAS:2012ae} and CMS \cite{Chatrchyan:2012tx} has attracted many considerations for new physics beside the higgs boson in standard model(SM) . One very interesting suggestion is \cite{Cheung:2011nv} which shows that Randall-Sundrum(RS) radion can be responsible for the excess. The physics behind RS model lies with the following geometry for the warped space-time \cite{Randall:1999ee},
\begin{equation}
ds^2=e^{-2kT(x)|\varphi|}[\eta_{\mu\nu}+G_{\mu\nu}(x)]dx^\mu dx^\nu +T^2(x)d\varphi^2 ,
\end{equation}
where $T(x)$ is referred to as the modulus field, $G_{\mu\nu}(x)$ as graviton and $k$ is a scale of the order of the (reduced) Planck scale $M_{pl}$. To explain the hierarchy problem, the compactification radius or the vacuum expectation value(vev) of the modulus field, $r_c\equiv\langle T(x)\rangle$, is required to satisfy the relation $kr_c\sim 12$.

The radion $\phi$ \cite{Goldberger:1999uk}, identified as the scalar bulk field to stabilize the modulus field, couples to SM particles as $
\mathcal{L}_{int}=\frac{\phi}{\Lambda_\phi}T^\mu{}_{\mu}$,
where $T_{\mu\nu}$ is energy-momentum tensor for SM particles and $\Lambda_\phi = \sqrt{6}M_{pl}e^{-kr_c\pi}$. This model leads to a larger branching ratio for $\phi\rightarrow gg \textrm{ or } \gamma\gamma$, relative to $h_{\textrm{SM}}\rightarrow gg \textrm{ or } \gamma\gamma$ in SM.
As shown in \cite{Cheung:2011nv}, with $\Lambda_\phi\sim 680$ GeV, the excess observed at the LHC can be explained by a $125$ GeV RS radion with $\sigma(H)Br(H\rightarrow \gamma\gamma)/\sigma Br_{\textrm{SM}}\sim 2.1$ and smaller values for other channels relative to the corresponding ones in SM.

The analysis above does not take into account of the constraint on other part of the RS model, namely the searches for a massive graviton at the LHC. In this note, we shall show that the results of LHC searches for graviton have interesting implications for the radion.

The $n$th massive Kaluza-Klein(KK) modes of $G_{\mu\nu}$ will also couple to SM particles, $
\mathcal{L}^{(n)}_{int}=\frac{1}{\Lambda_{G}}G^{(n)}_{\mu\nu}T^{\mu\nu}$,
where $\Lambda_{G}=M_{pl}e^{-kr_c\pi}$. The mass of the $n$th KK graviton is given by $M_{G_n}=kx_ne^{-kr_c\pi}=x_n\frac{k}{M_{pl}}\Lambda_{G}$, where $x_n$ is the $n$th solution of $J_1(x_n)=0$, and $J_1$ is the Bessel function. In the following, we will focus on the first KK mode with $x_1=3.83$, $M_{G}\equiv M_{G_1}$.

The couplings of the first KK graviton with SM particles are proportional to $1/\Lambda_{G}$ or $x_1k/M_{pl}$ for a fixed $M_G$. Limits put on $M_{G}$ for specified $k/M_{pl}$ can then be transferred to limits on $\Lambda_{G}$, therefore on $\Lambda_{\phi}$ due to the relation, $\Lambda_{\phi}=\sqrt{6}\Lambda_{G}$. For example, using dijet final states, CMS \cite{Chatrchyan:2011ns} with $1 \textrm{ fb}^{-1}$ has excluded a RS graviton mass below $1$ TeV for $k/M_{pl}=0.1$. A straightforward calculation gives that the corresponding $\Lambda_{\phi}=\frac{\sqrt{6}}{x_1 k/M_{pl}}M_G=6.4$ TeV. More recently, using dilepton final states, ATLAS \cite{ATLASdilepton} with $5 \textrm{ fb}^{-1}$ show that a RS graviton mass below $2.16$ TeV is excluded at $95\%$ confidence level also with $k/M_{pl}=0.1$, then the corresponding $\Lambda_{\phi}=13.8$ TeV.

\begin{figure}[htb]
\includegraphics[width=0.42\textwidth,height=2.0in]{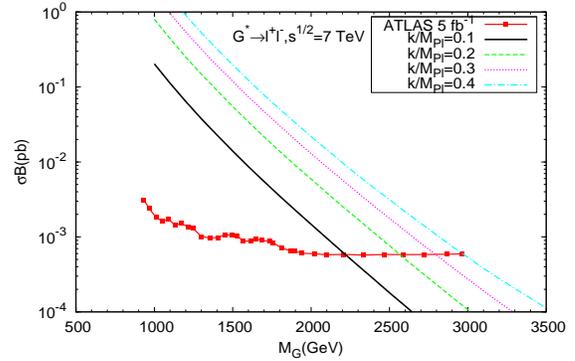}
\caption{Limit on the mass of RS graviton for various $k/M_{pl}$, where the box points are taken from \cite{ATLASdilepton}.}
\label{fig:Gll}
\end{figure}

A smaller value of $\Lambda_{\phi}$ then requires a larger $k/M_{pl}$, although the latter of order $0.1$ or less is preferred theoretically\cite{Davoudiasl:2000wi}. However, a larger $k/M_{pl}$ means a more stringent constraint on $M_{G}$ because the cross section for the graviton's production at the LHC is proportional to $(k/M_{pl})^2$. As shown in Fig.~\ref{fig:Gll}, when $k/M_{pl}=0.3$, the limit for $M_G$ is $2.8$ TeV, then we have $\Lambda_{\phi}=5.97$ TeV.
\begin{figure}[t]
\includegraphics[width=0.40\textwidth,height=2.0in]{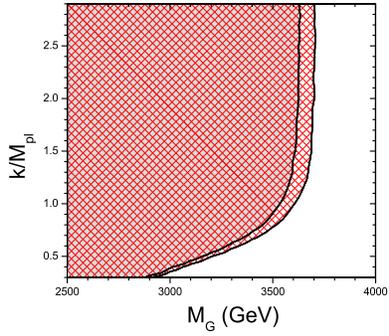}
\caption{Conservative limit on $k/M_{pl}$ and $M_G$, where the shadowed region is excluded at least at $95\%$ confidence level and the region between two solid lines indicates the effect of $10\%$ uncertainty.}
\label{fig:Contour}
\end{figure}

One may want to extend to even larger $k/M_{pl}$ and hope to accommodate $\Lambda_\phi=680$ GeV. The obstacle is that there is an upper limit for $k/M_{pl}$ theoretically given by \cite{Agashe:2007zd}, $k/M_{pl}<\sqrt{3\pi^3/5\sqrt{5}}\simeq 2.88$. We show the conservative constraint for large $k/M_{pl}$ in Fig.~\ref{fig:Contour}. A limit of $M_G=3.5$ TeV will give $\Lambda_{\phi}=2.24$ TeV for $k/M_{pl}\simeq 1$. Even the largest but highly theoretically disfavoured $k/M_{pl}\simeq 2.88$ results in $\Lambda_{\phi}=0.8$ TeV and $\sigma(H)Br(H\rightarrow \gamma\gamma)/\sigma Br_{\textrm{SM}}\sim 1.5$.

In summary, it is unlikely to have a $125$ GeV RS radion with $\Lambda_\phi=680$GeV and accommodate with both experimental and theoretically constraints.


\begin{thebibliography}{References}
\bibitem{ATLAS:2012ae}
  G.~Aad {\it et al.}  [ATLAS Collaboration],
  Phys.\ Lett.\ B {\bf 710}, 49 (2012)
  [arXiv:1202.1408 [hep-ex]].

\bibitem{Chatrchyan:2012tx}
  S.~Chatrchyan {\it et al.}  [CMS Collaboration],
  arXiv:1202.1488 [hep-ex].

\bibitem{Cheung:2011nv}
  K.~Cheung and T.~-C.~Yuan,
  Phys.\  Rev.\  Lett.\  108, {\bf 141602} (2012)
  [arXiv:1112.4146 [hep-ph]].

\bibitem{Randall:1999ee}
  L.~Randall and R.~Sundrum,
  Phys.\ Rev.\ Lett.\  {\bf 83}, 3370 (1999)
  [hep-ph/9905221].

\bibitem{Goldberger:1999uk}
  W.~D.~Goldberger and M.~B.~Wise,
  Phys.\ Rev.\ Lett.\  {\bf 83}, 4922 (1999)
  [hep-ph/9907447].

\bibitem{Chatrchyan:2011ns}
  S.~Chatrchyan {\it et al.}  [CMS Collaboration],
  Phys.\ Lett.\ B {\bf 704}, 123 (2011)
  [arXiv:1107.4771 [hep-ex]].

\bibitem{ATLASdilepton}
 The ATLAS Collaboration, ATLAS-CONF-2012-007.

\bibitem{Davoudiasl:2000wi}
  H.~Davoudiasl, J.~L.~Hewett and T.~G.~Rizzo,
  Phys.\ Rev.\ D {\bf 63}, 075004 (2001)
  [hep-ph/0006041].

\bibitem{Agashe:2007zd}
  K.~Agashe, H.~Davoudiasl, G.~Perez and A.~Soni,
  Phys.\ Rev.\ D {\bf 76}, 036006 (2007),
  B.~Grzadkowski and J.~F.~Gunion,
  arXiv:1202.5017 [hep-ph].


\end{thebibliography}
\end{document}